\documentclass[prl,aps,twocolumn,superscriptaddress,preprintnumbers,floatfix]{revtex4-1}
\pdfoutput=1
\usepackage{amsmath,amssymb,bm,graphicx,bbold,epsf,colordvi,bbm}
\usepackage[small]{subfigure}
\usepackage{xspace,cancel}

\allowdisplaybreaks 
\newcommand\slurp[1]{#1}
\newcommand\sslurp[2]{#1{#2}}
{\catcode`/=\active \expandafter}%
\slurp{\newcommand/}{/\penalty1000\hskip0pt\relax}
{\catcode`:=\active \expandafter}%
\slurp{\newcommand:}{:\penalty1000\hskip0pt\relax}

\newcommand\addspace{\ifcat\nextchar a\spacefactor999. \else.\fi}
{\catcode`\.=\active \expandafter}%
\slurp{\newcommand.}{\futurelet\nextchar\addspace}

\usepackage[unicode]{hyperref} 

\ifx\href\undefined\def\href#1{}\fi
\ifx\texorpdfstring\undefined\def\texorpdfstring#1#2{#1}\fi

\newcommand\myslash{/} 
\newcommand{\be}{\begin{equation}}
\newcommand{\ee}{\end{equation}}
\newcommand{\nn}{\nonumber}

\def\d{{\rm d}}
\def\OMIT#1{{}}

\newcommand{\Dslash}{D\hspace{-0.25cm}\myslash \, }
\newcommand{\id}{\mathbbm{1}}

\newcommand{\e}{\mathrm{e}}

\newcommand{\eq}[1]{Eq.~\eqref{#1}}

\newcommand{\bn}{\bar n}

\begin{document}

%%%%%%%%%%%%%%%%%%%%%%%%%%%%%%%%%%%%%%%%%%%%%%%%%%%%%%%%%%%%%%%%%%%%%%%%%%%%%%%%
% Title page
%%%%%%%%%%%%%%%%%%%%%%%%%%%%%%%%%%%%%%%%%%%%%%%%%%%%%%%%%%%%%%%%%%%%%%%%%%%%%%%%
\preprint{\vbox{\hbox{ACFI-T14-18}}}
\preprint{\vbox{
		\hbox{MIT--CTP 4593}
	}}

%\begin{flushright}
%MZ-TH/12-13\\
% arXiv:1203.nnnn
% v0: March 22, 2012
% v1: March 30, 2012
%March 30, 2012 
%end{flushright}

\title{Heavy Dark Matter Annihilation from Effective Field Theory}

\author{Grigory Ovanesyan}
\affiliation{Physics Department, University of Massachusetts Amherst, Amherst, Massachusetts 01003, USA}

\author{Tracy R.~Slatyer}
\affiliation{Center for Theoretical Physics, Massachusetts Institute of Technology, Cambridge, MA 02139, USA\vspace{0.5ex}}

\author{Iain W.~Stewart\vspace{1.ex}}
\affiliation{Center for Theoretical Physics, Massachusetts Institute of Technology, Cambridge, MA 02139, USA\vspace{0.5ex}}

\begin{abstract}

We formulate an effective field theory description for SU(2)$_L$ triplet fermionic dark matter by combining nonrelativistic dark matter with gauge bosons in the soft-collinear effective theory. For a given dark matter mass, the annihilation cross section to line photons is obtained with 5\% precision by simultaneously including Sommerfeld enhancement and the resummation of electroweak Sudakov logarithms at next-to-next-to-leading logarithmic order.  Using these results, we present more accurate and precise predictions for the gamma-ray line signal from annihilation, updating both existing constraints and the reach of future experiments.

\end{abstract}

\maketitle

%{\bf Introduction.}\label{sec:introduction}\ \ 
%
If Weakly Interacting Massive Particles (WIMPs) exist at the TeV scale, their annihilations in the present day -- and hence their signatures in indirect dark matter (DM) searches -- experience large corrections that are not well described by a simple perturbative expansion in the coupling. On the one hand, exchanges of electroweak gauge bosons and photons between DM particles in the initial state give rise to a long-range potential. This effect, known as the ``Sommerfeld enhancement'', sums large corrections $\sim \sum_k( \alpha_2 m_\chi/m_W)^k$ and has been extensively studied in the literature (e.g. \cite{Hisano:2003ec, Hisano:2004ds, Cirelli:2007xd, ArkaniHamed:2008qn}). 
On the other hand, the large hierarchy between the DM mass $m_\chi$ and gauge boson mass $m_W$ generates large logarithmic corrections, and this 
has not yet been explored in detail. A study~\cite{Hryczuk:2011vi} of wino DM annihilation at one-loop found $\mathcal{O}(1)$ corrections, which change the predicted annihilation cross section by a factor of a few. This is a signal of large logarithmic corrections, $\sim \sum_k (\alpha_2 \ln^{2,1} m_\chi^2/m_W^2)^k$, whose resummation  is the focus of this work.  

This goal is not an abstract one: existing ground-based gamma-ray telescopes can probe the annihilation of multi-TeV DM \cite{Cohen:2013ama,Fan:2013faa}, and future colliders may also have sensitivity \cite{Cirelli:2014dsa}. Null results from the LHC already place stringent lower bounds on the SUSY spectrum; so, while direct constraints on DM from the LHC are still not especially strong, the lack of a detection of new physics below the TeV scale motivates consideration of heavier-than-TeV DM and its properties. As one example, models of ``split supersymmetry'' \cite{Giudice:2004tc, ArkaniHamed:2004fb} can preserve the unification of gauge couplings with fermionic superpartners at the TeV scale \cite{Arvanitaki:2012ps}. It is therefore imperative to understand how to translate models of heavy DM into signal predictions with accurate theoretical cross sections.

We focus on pure wino DM and its annihilation to line gamma rays, $\chi^0 \chi^0 \rightarrow \gamma \gamma$, $\gamma Z$. At the weak scale and above such spectral lines have zero astrophysical background, so detection would be a smoking gun for new physics. We show that Sommerfeld enhancement effects can be factorized from large logs to all orders in $\alpha_2$, and compute the cross section at next-to-next-to-leading logarithmic (NLL) order for line photon production, including an estimate of theoretical uncertainties.

{\bf Dark matter model.}
We do not know yet what the non-gravitational interactions of Dark Matter (DM) are.  Here we are interested in DM being an SU(2)$_L$ triplet of Majorana fermions, a scenario under active investigation~\cite{Hryczuk:2011vi,Ciafaloni:2012gs,Cohen:2013ama,Cirelli:2014dsa} both in the context of the SUSY wino and more generally. The DM triplet can be written as: 

% a 2-by-2 matrix \cite{Ciafaloni:2012gs}\\
\begin{eqnarray}
\chi=\left( \begin{array}{cc}
\chi^0/\sqrt{2} & \chi^+  \\
\chi^- & -\chi^0/\sqrt{2} \end{array} \right),
\end{eqnarray}
which transforms from left and right under the $SU(2)_L$ gauge group of Standard Model (SM). We extend the SM Lagrangian by including
%\begin{eqnarray}
$\mathcal{L}_{\rm DM} = \frac{1}{2}\text{Tr}\,\bar{\chi} \big(\,i \Dslash\,- M_{\chi} \big) \chi$,
%-\frac{1}{2}\,\text{Tr}\,\bar{\chi}M_{\chi}\chi,
%\end{eqnarray}
where the trace sums over the $SU(2)_L$ indexes and the covariant derivative couples the DM to SM gauge bosons $W^{1,2,3}$ or equivalently $\gamma, W,Z$ ($\chi$ has zero hypercharge). 
In principle the mass mixing and splitting can be described by an arbitrary matrix $M_{\chi}$; however in the minimal scenario it is $M_\chi = m_{\chi} \id $. A small mass splitting between $\chi^0$ and $\chi^-$ is generated radiatively; and we take it to be $\delta=0.17$ GeV for the Sommerfeld calculation, but ignore it in the Sudakov calculation. The presence of this splitting means the $\chi^0$ constitutes all the stable DM. However, initial-state exchange of W bosons allows excitation from a $\chi^0 \chi^0$ two-body state into an (off-shell) $\chi^+ \chi^-$ state, and in calculating the Sommerfeld-enhanced cross section the matrix elements for annihilation from $\chi^0 \chi^0$ and $\chi^+ \chi^-$ initial states must therefore be included.

We focus on the late-time annihilation of triplet DM and thus assume there are no on-shell $\chi^\pm$ present in the DM halo. We also assume $s$-wave annihilation, since $p$-wave and higher terms are suppressed by at least the square of the small DM velocity in the local halo ($v \sim 10^{-3}$). This ensures the $\chi^0\chi^0$ initial state is a spin singlet. This also implies that annihilation to three gauge bosons is forbidden by CP conservation \cite{Hryczuk:2011vi}, so we consider only two-body final states.

{\bf Electroweak corrections in NRDM-SCET.}
The soft-collinear effective theory (SCET)~\cite{Bauer:2000ew,Bauer:2000yr,Bauer:2001ct,Bauer:2001yt} has been used to describe electroweak radiative corrections in high-energy processes via exchanges of weak gauge bosons of the SM gauge group~\cite{Chiu:2009mg,Chiu:2009ft}. We generalize this formalism to the case with heavy nonrelativistic dark matter (NRDM) in the initial state, and use it to calculate $\chi\chi \to ZZ, Z\gamma,\gamma\gamma$. The calculation can be broken into pieces: constructing operators, matching at a high scale $\mu\simeq 2m_\chi$, running down to $\mu\simeq m_Z$, and calculating matrix elements at this low scale which include the Sommerfeld enhancement.

{\it EFT and High scale matching.}
At the high scale $\mu_{m_{\chi}}\simeq \sqrt{s}=2m_{\chi}$ we match the annihilation process in the full theory ${\cal L}_{\rm SM}+{\cal L}_{\rm DM}$   onto a set of leading order operators $O_r$ in our effective theory NRDM-SCET:
\begin{align}
 {\cal L}_{\rm ann}^{(0)} = \mbox{$\sum_{r=1}^2$} \: C_r(m_\chi,\mu)\: O_r(m_{W/Z},v,\mu) 
  \,.
\end{align}
There are only two operators in the complete basis for spin-singlet S-wave annihilation of DM: 
\begin{align}  \label{eq:Or}
 O_{r} &=  \big(\chi_v^{aT} i\sigma_2 \chi_v^b\big) \: \big(S_{r}^{abcd} \: {\cal B}_{n\perp}^{ic} {\cal B}_{\bn\perp}^{jd}\big) i \epsilon^{ijk} (n-\bn)^k  \,, \nn\\
  S_1^{abcd} &= \delta^{ab} ({\cal S}_n^{ce} {\cal S}_{\bn}^{de}) 
    \,,\
  S_2^{abcd} = ({\cal S}_v^{ae} {\cal S}_n^{ce}) ({\cal S}_v^{bf} {\cal S}_{\bn}^{df}) 
  \,.
\end{align}
Here $v=(1,0,0,0)$,  $n=(1,\hat n)$, and $\bn=(1,-\hat n)$ with $\hat n$ the direction of an outgoing gauge boson.  $\chi_v^a$ is a non-relativistic two-component fermion DM field in the adjoint representation, ${\cal B}_{n,\bn}$ contain the observed (collinear) gauge bosons, and the ${\cal S}_\kappa={\cal S}_\kappa[\kappa\cdot A_s]$ are adjoint Wilson lines of soft gauge bosons along the $\kappa=n,\bn,v$ directions. Without soft gluons
there are only two possible contractions of gauge indices, $\delta^{ab}\delta^{cd}$ and $\delta^{ac}\delta^{bd}$, since $(\chi_v^{aT} i\sigma_2 \chi_v^b) = \chi_v^{a\alpha} \chi_v^{b\beta} \epsilon_{\alpha\beta}$ is symmetric in $(ab)$. Due to the factorization properties of soft gluons for heavy particles $v$, or collinear particles $n$, $\bn$, the addition of the soft ${\cal S}_\kappa$ Wilson lines does not change this, see~\cite{Bauer:2001yt}. The final state gauge bosons are also in a spin-singlet with orthogonal polarizations so they must be contracted with $\epsilon^{ijk}$. 
The outgoing energetic gauge bosons appear in the adjoint collinear gauge invariant building block
${\cal B}_{n\perp}^{\mu a}= i/(i\bn\cdot\partial_n) \bn_\nu G_n^{\nu \mu b} {\cal W}_n^{ba} = A_{n\perp}^{\mu a} - \frac{k_\perp^\mu}{\bn\cdot k} \bn\cdot A_{n}^a + \ldots $,
where $A_n^{\mu a}$ is the $n$-collinear gauge boson field, and ${\cal W}_n^{ba}={\cal W}_n^{ba}[\bn\cdot A_n]$ is a collinear Wilson line in the adjoint representation. For the definition of ${\cal B}_{\bn\perp}^{\mu a}$ simply swap $n\leftrightarrow \bn$. In addition to the hard annihilation process encoded in ${\cal L}_{\rm ann}^{(0)}$, we will also use the leading order SCET${}_{\rm II}$ Lagrangian ${\cal L}_{\rm SCET}^{(0)}$ and leading order nonrelativistic Lagrangian for DM ${\cal L}_{\rm NRDM}^{(0)} =  \chi_v^\dagger (i v\cdot \partial +\vec{\nabla}^2/2m_\chi) \chi_v + \hat V[\chi_v^{(\dagger)}](m_{W,Z})$, where $\hat V$ is an operator giving the Yukawa and Coulombic potentials from potential exchange of the $W, Z, \gamma$. 

To determine the Wilson coefficients $C_r$ at the high scale we match from the full theory onto the effective theory. Since $C_r$ only contain ultraviolet physics this matching can be done in the unbroken SM with $m_{W}=m_Z=0$.  At tree level we find $C_1(\mu_{m_\chi}) = - C_2(\mu_{m_\chi})= -\pi \alpha_2(\mu_{m_\chi}) /m_\chi$, 
where $\alpha_2 = g^2/4\pi = \alpha /\sin^2\bar \theta_W$.

{\it Sommerfeld-Sudakov Factorization}
Since ${\cal L}_{\rm NRDM}^{(0)}$ contains no interactions with soft or collinear gauge bosons, and ${\cal L}_{\rm SCET}^{(0)}$ contains no interactions with $\chi_v$s, the matrix element for the $\chi^0\chi^0$ evolution and annihilation factorizes from the matrix element involving the final state $X$:
\begin{align}  \label{eq:factorization}
  C_r\langle X |  O_r | \chi^0 \chi^0 \rangle  
 &= \!\big[ C_r\, i\epsilon^{ijk} (n\!-\!\bn)^k \langle X |S_{r}^{abcd} \,
   {\cal B}_{n\perp}^{ic} {\cal B}_{\bn\perp}^{jd}\big)  | 0 \rangle \big]
   \nn\\
 & \times \langle 0 | \chi_v^{aT} i\sigma_2 \chi_v^b | \chi^0 \chi^0 \rangle
 .
\end{align}
For the spin-singlet state $|(\chi^a\chi^b)_S\rangle = \epsilon^{\beta\alpha} |\chi_\alpha^a \chi_\beta^b \rangle / \sqrt{2}$,
the Sommerfeld enhancement factors are encoded in
\begin{align} \label{eq:sdefn}
 &  \big\langle 0 \big| \chi_v^{3T} i\sigma_2 \chi_v^3 \big| (\chi^0 \chi^0)_S \big\rangle = 4 \sqrt{2} m_\chi s_{00} \,, \\
 & \big\langle 0 \big| \chi_v^{+T} i\sigma_2 \chi_v^- \big| (\chi^0 \chi^0)_S \big\rangle= 4 m_\chi s_{0\pm} \,,\nn
\end{align}
where the matrix elements are evaluated using the potential $\hat V$. For these channels the corresponding matrix elements on the first line of (\ref{eq:factorization}) can be denoted $F_{0}^X$ and $F_{\pm}^X$, thus giving an all-orders factorized result for the spin-singlet annihilation amplitudes
\begin{align} 
 \mathcal{M}_{\chi^0 \chi^0\rightarrow X} &=4 m_\chi 
  \big(\sqrt{2} s_{00} F_0^X + s_{0\pm} F_\pm^X \big) \,,
  \\
 \mathcal{M}_{\chi^+ \chi^-\rightarrow X} &=2\sqrt{2} m_\chi 
  \big(\sqrt{2} s_{\pm 0} F_0^X +  s_{\pm\pm} F_\pm^X \big)
  \,. \nn
\end{align}
In the one-loop calculation of~\cite{Hryczuk:2011vi}, the coefficients $s_0^{\text{\cite{Hryczuk:2011vi}}}=s_{00}$ and $s_{\pm}^{\text{\cite{Hryczuk:2011vi}}}=s_{0\pm}$ were also included as multiplicative factors, which is consistent with this factorization. 
We obtain the Sommerfeld coefficients $s_{00}$ and $s_{0\pm}$ by solving the Schroedinger equation numerically (see e.g. Appendix A of \cite{Cohen:2013ama} for details). Note that at tree level $s_{00}=s_{\pm\pm}=1$ and $s_{0\pm}=s_{\pm0}=0$. 

With SU(2)$_L$ symmetry the gauge index structure of the first line of (\ref{eq:factorization}) implies that the SCET perturbative corrections at any order are encoded in just two Sudakov form factors, $\Sigma_1$ and $\Sigma_2$. The gauge boson masses induce symmetry breaking corrections at NLL which are included by using $\Sigma_{1,2}^W$ for the $W^+W^-$ final state, so
\begin{align}
  F_0^{\gamma\gamma}\! & = P_{\gamma\gamma} (\Sigma_1 \!-\! \Sigma_2) \,,
 & F_{\pm}^{\gamma\gamma} & = 2 P_{\gamma\gamma} \Sigma_1 \,, 
  \\ 
  F_0^{W^+W^-}\! &= P_{W} \Sigma_1^W \,,
 & F_{\pm}^{W^+W^-} &= P_{W} (2\Sigma_1^W \!-\! \Sigma_2^W) \,,
 \nn
\end{align} 
where the prefactors are $P_{\gamma\gamma} = -e^2 \epsilon_{n\perp}^i \epsilon_{\bn\perp}^j \epsilon^{ijk} \hat n^k/(2m_\chi)$ and $P_W = (g^2/e^2) P_\gamma$. For $F_0^{\gamma Z}$ and $F_0^{ZZ}$ one simply replaces $P_{\gamma\gamma}$ by $P_{\gamma Z}= \cot\bar\theta_W P_{\gamma\gamma}$ or $P_{ZZ}= \cot^2\bar\theta_W P_{\gamma\gamma}$.  At tree level the form factors are all unity, $\Sigma_1=\Sigma_2=1$. 

For the $\gamma \gamma$ and $\gamma Z$ final states there is no tree-level annihilation from $\chi^0 \chi^0$, so we normalize by writing
\begin{align}
  \sigma_{\chi^0 \chi^0 \rightarrow X} = \sigma^{\text{tree}}_{\chi^+ \chi^- \rightarrow X} \big| s_{00} (\Sigma_1 - \Sigma_2) +\, \sqrt{2} s_{0 \pm} \Sigma_1 \big|^2. \label{eq:sommerfeldxsec}
\end{align}

{\bf Sudakov Resummation.}
We now calculate the Sudakov form factors $\Sigma_{1,2}$. For simplicity, in this calculation we take all DM components to have a common mass $m_{\chi}$. The operators $O_{1,2}$ in (\ref{eq:Or}) mix under renormalization and the resummation of $\alpha_2\ln^{2,1}(m_\chi^2/m_W^2)$ corrections is achieved by 
finding their SCET anomalous dimension matrix, and running between the high scale $\mu_{m_{\chi}}\simeq 2m_\chi$ and the low scale $\mu_{Z}\simeq m_Z$.  For NLL order resummation we need the two-loop cusp and one-loop non-cusp  anomalous dimensions, plus the high scale matching  at tree level.
The one-loop anomalous dimension matrix for an operator with standard model quantum numbers and any number of single collinear building blocks was derived in Ref. \cite{Chiu:2009mg}, and we will make use of their results, including the $\Delta$-regulator \cite{Chiu:2009yx,Chiu:2009mg}. Our case differs from this general result because the incoming nonrelativistic DM fields are in the same direction $v$, and hence we have two soft $S_v$ Wilson lines that can interact with each other or self-interact.
 
The anomalous dimension matrix for $(C_{1}\ C_2)^T$ is
\begin{eqnarray}
&& \hat\gamma= 2 \gamma_{W_T} \id + \hat\gamma_S \,.
\end{eqnarray}
Here $\gamma_{W_T}$ is the collinear anomalous dimension of ${\cal B}_{n\perp}^{ia}$ which only mixes into itself, and hence multiplies a diagonal matrix. Including the two-loop cusp and one-loop non-cusp terms it is equal to~\cite{Chiu:2009mg}:
\begin{align} \label{eq:anomdimWT}
\gamma_{W_T}^{\rm NLL}=\frac{\alpha_2}{4\pi} \Gamma_0^g \ln\frac{2m_\chi}{\mu} -\frac{\alpha_2}{4\pi} b_0
  + \Big(\frac{\alpha_2}{4\pi}\Big)^2 \Gamma_1^g \ln\frac{2m_\chi}{\mu}
  ,
\end{align}
where here and below $\alpha_2(\mu)$ is in the $\overline{\rm MS}$ scheme, and for $SU(2)$ in the SM, $C_A=2$, $b_0=19/6$ is the one-loop $\beta$-function, the cusp anomalous dimensions are $\Gamma_0^g=4 C_A=8 $ and $\Gamma_1^g = 8\left(\frac{70}{9}-\frac{2}{3}\pi^2\right)$. When integrating, we will also need the two-loop $\beta$-function $b_1=-{35}/{6}$.

\begin{figure}[t!]  
  \centering
    \includegraphics[width=0.47\textwidth]{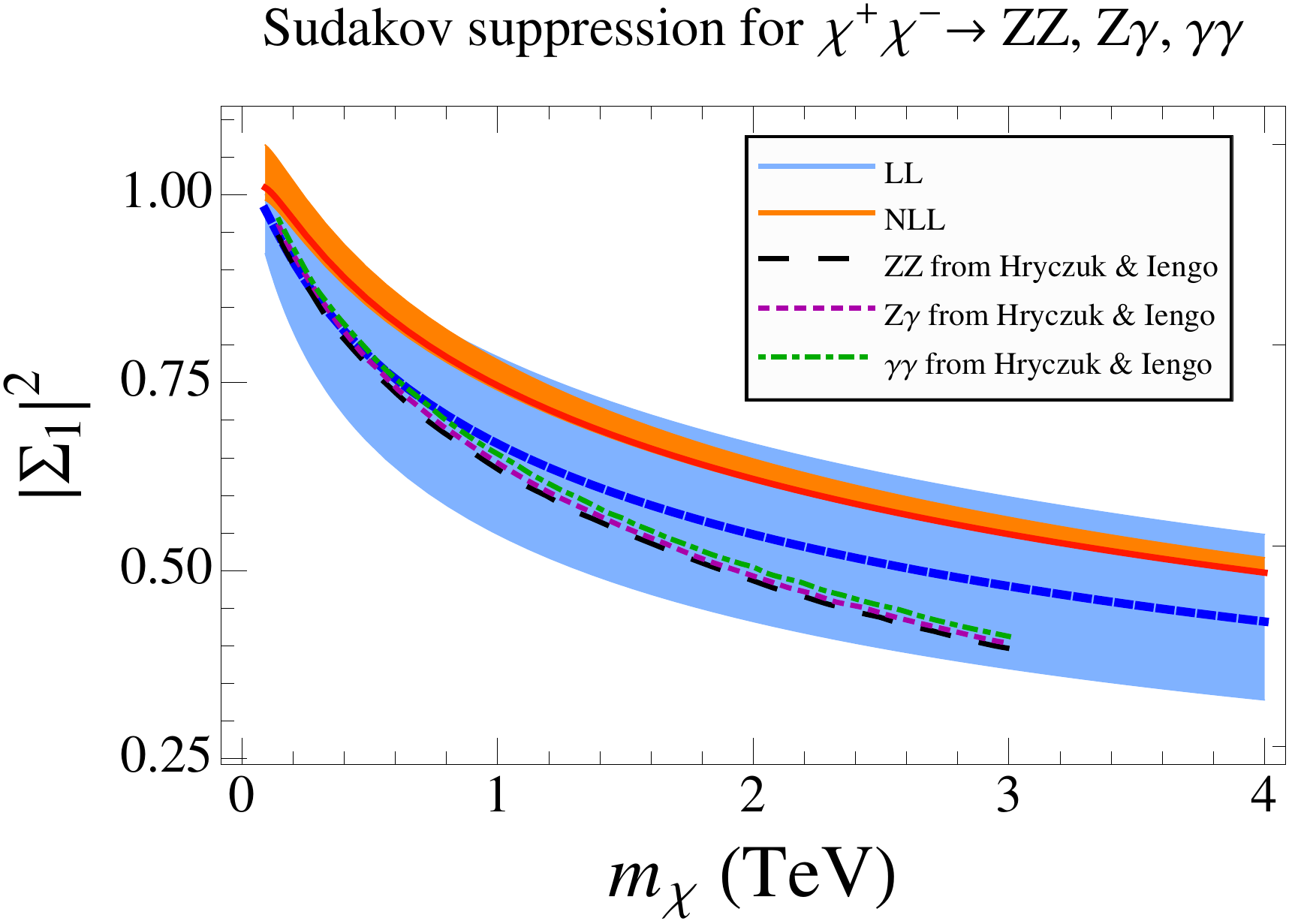}\qquad\includegraphics[width=0.47\textwidth]{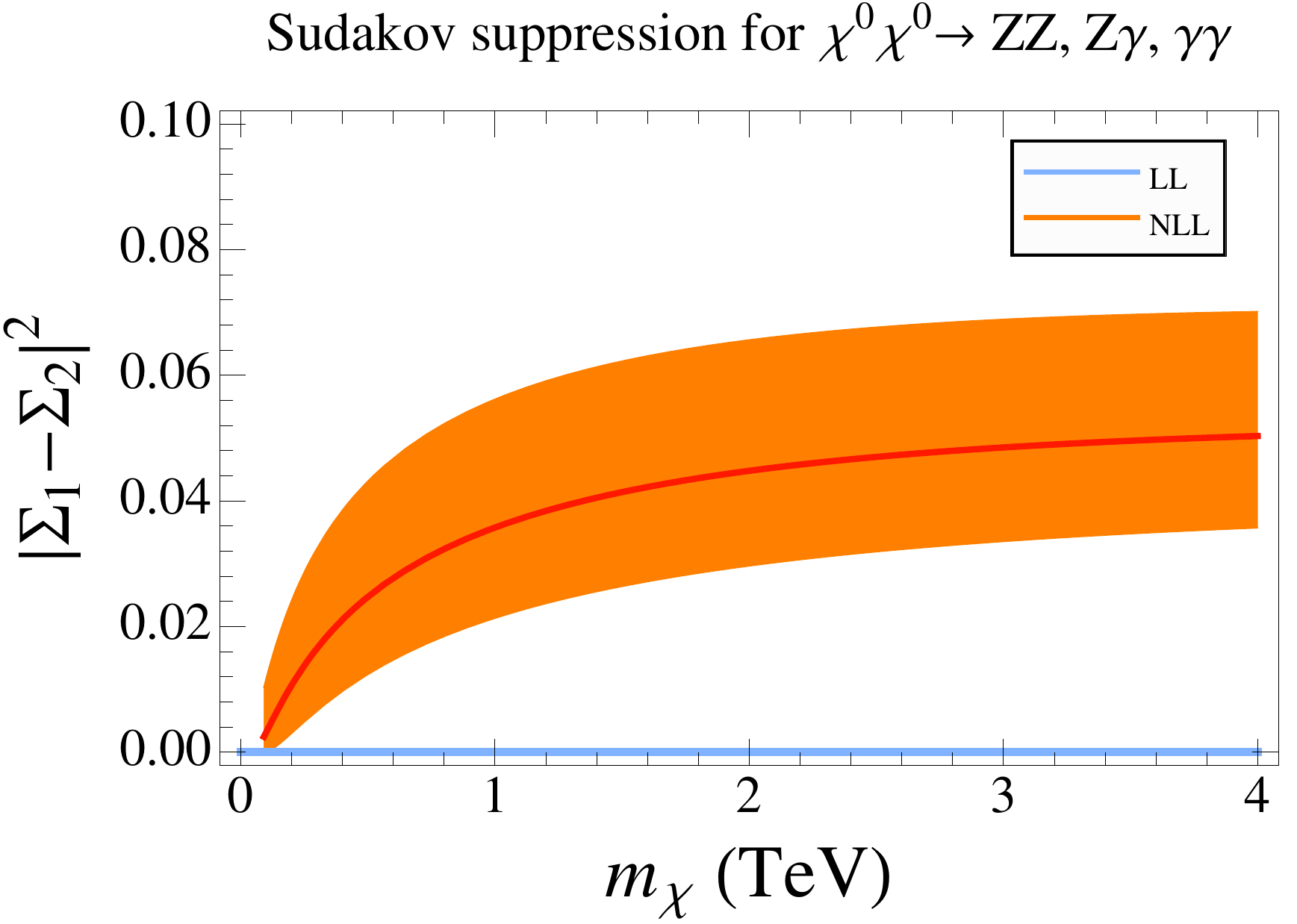}
    \vspace{-0.2cm}
    \caption{Resummed leading and next-to-leading logarithmic electroweak corrections for  $\chi^+\chi^-, \chi^0\chi^0\rightarrow ZZ, Z\gamma, \gamma\gamma$. Only high scale variation by a factor of 2 from $\mu_{m_{\chi}}=2m_{\chi}$ is shown. Low scale variation has a 20\% smaller error band for the top plot and a 5\% bigger error band for the bottom plot.   }\label{figNLL}
\end{figure}
 
The soft anomalous dimension $\hat\gamma_S$ encodes the running and mixing of the soft factors $S_{1,2}^{abcd}$ and hence has non-trivial structure.
After canceling the regulator dependent part with the zero-bin subtracted \cite{Manohar:2006nz} collinear graphs, the non-zero one-loop contributions come from:  wavefunction renormalization from self contracting a $S_v$, connecting the two $S_v$ Wilson lines, and connecting the $S_n$ and $S_{\bn}$ Wilson lines. The wavefunction renormalization is the same as HQET, $\gamma_{h_v} = - C_A \alpha_2 / (2\pi)$. 
The full result needed at NLL is
\begin{eqnarray}
\hat \gamma_{S}^{\rm NLL}=\frac{\alpha_2}{\pi}(1-i\pi)
\bigg( \begin{array}{ccc}
2 & \,\,\,\,\,\,\,\,1 \\
0 & \,\,\,\,-1  \end{array} \bigg)
 - \frac{2\alpha_2}{\pi} 
\bigg( \begin{array}{ccc}
1 & \,\,\,\,\,\,0 \\
0 & \,\,\,\,\,\,1  \end{array} \bigg)
 \,.
\end{eqnarray}

At the low scale $\mu_{Z}\simeq m_{Z}$ the operators $O_1,O_2$ are matched onto a operators with $W,Z,\gamma$s, and effects associated with the gauge boson masses are included from low scale matching (or using the rapidity renormalization group~\cite{Chiu:2012ir,Becher:2010tm}).  Here we are interested in neutral transverse final state gauge bosons, where the matching at NLL order reads \cite{Chiu:2009ft}
${\cal B}^3_{\perp}\rightarrow \exp(D)\left(Z_{\perp}\,\cos\theta_W+A_{\perp}\,\sin\theta_W\right)$
with
\begin{eqnarray}
&& D(\mu_Z)=\frac{\alpha_2(\mu_{Z})}{2\pi}\ln\frac{4m_\chi^2}{\mu_{Z}^2}\ln\frac{m_W^2}{\mu_{Z}^2}\,.\label{eq:lowscale2}
\end{eqnarray}
The Sommerfeld enhancement factors in (\ref{eq:sdefn}) are low scale matrix elements which are also calculated at $\mu_{Z}\simeq m_Z$. 
\begin{figure*}[t!]
%\begin{center}
\includegraphics[width=0.48\textwidth]{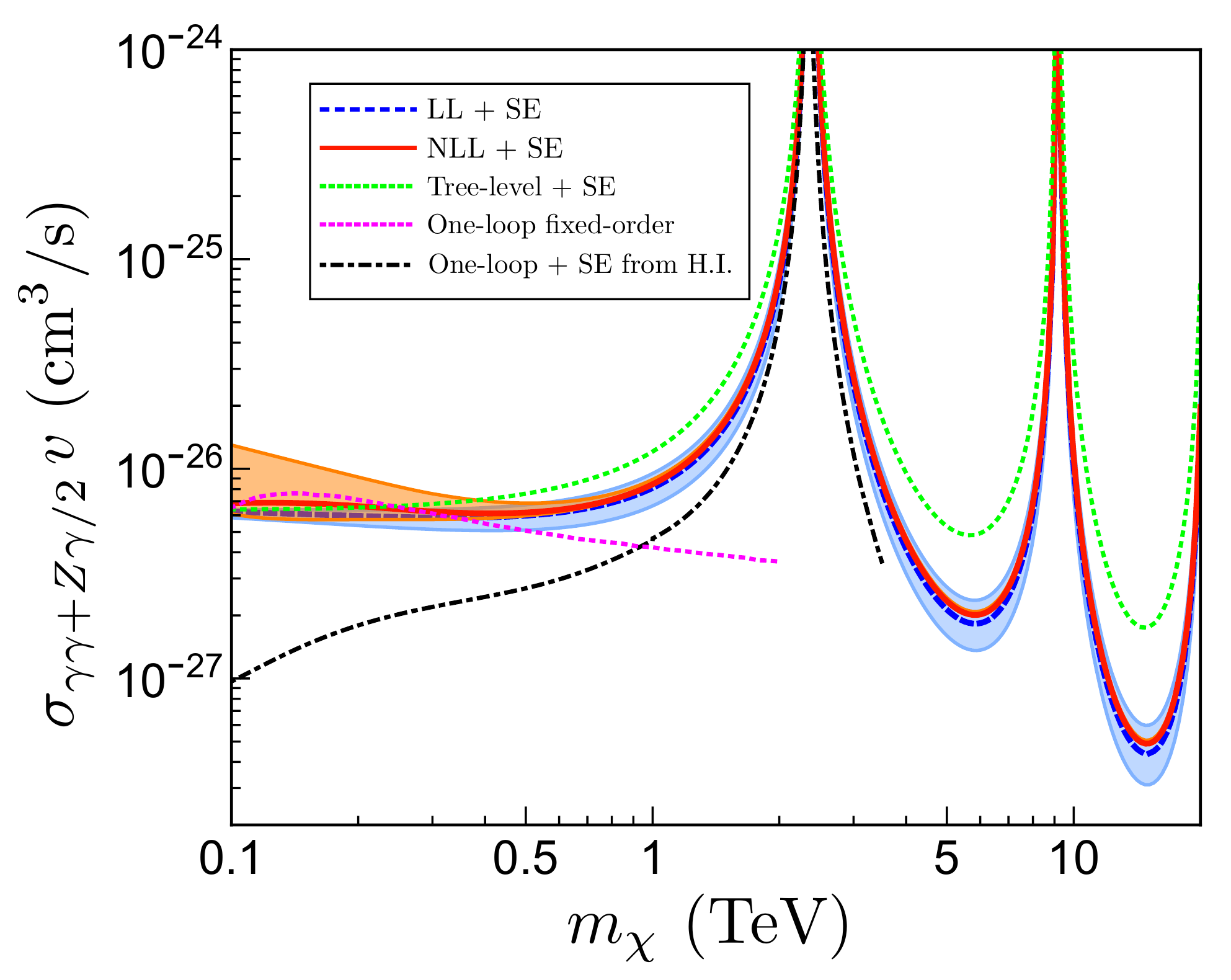}\hspace{0.03\textwidth}
\includegraphics[width=0.48\textwidth]{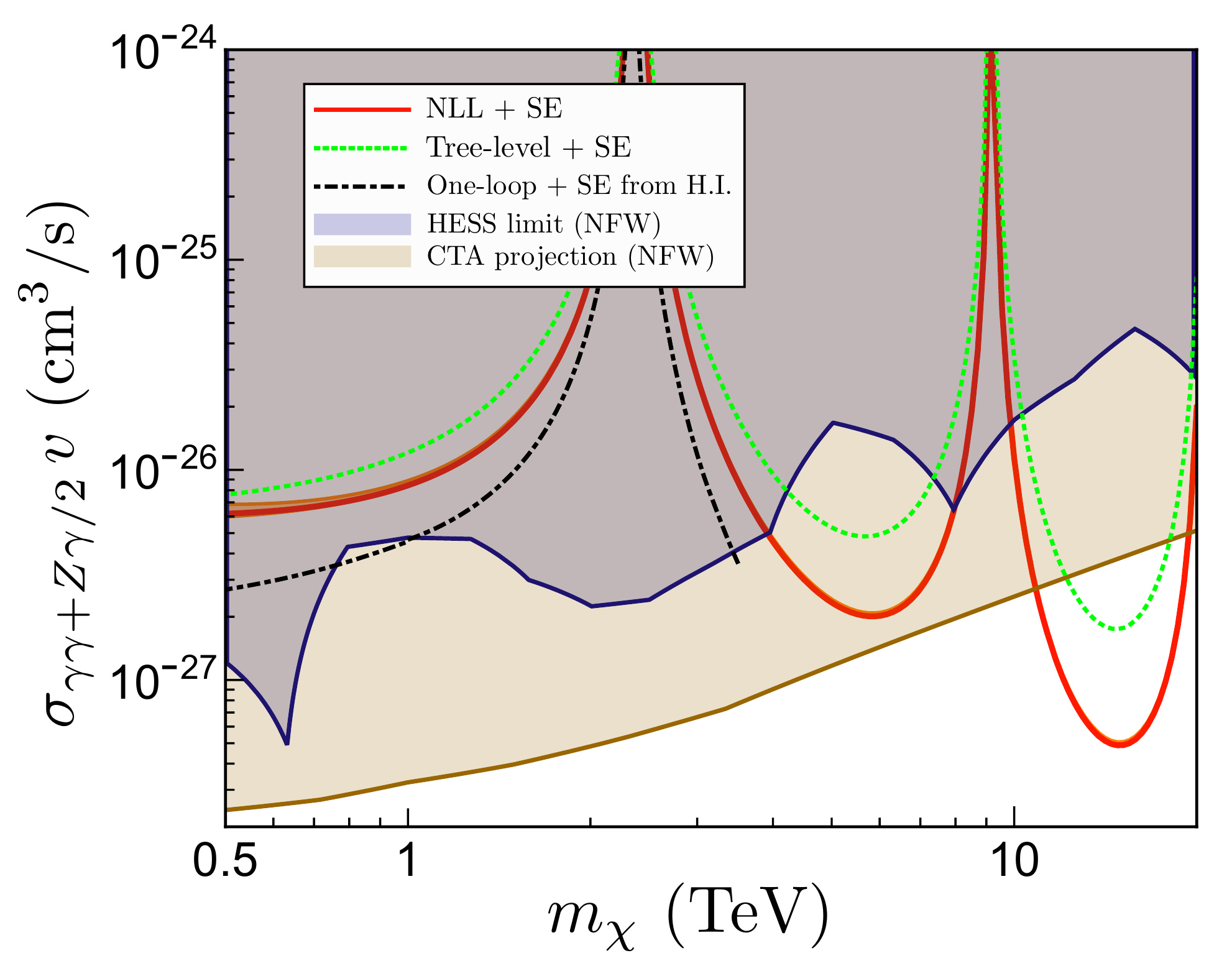}\vspace{-.4cm}
%\end{center}
\caption{Left panel: Our NLL+SE cross section for $\chi^0 \chi^0$ annihilation to line photons from $\gamma \gamma$ and $\gamma Z$, compared to earlier results. Right panel: current bounds from H.E.S.S and projected reach of 5 hours of CTA observation time, overlaid with our (and previous) cross section predictions, for an NFW profile.}\label{figxsec}
\end{figure*}

{\it Analytical resummation formula at NLL order.}
The resummed amplitude is 
\begin{align}\label{eq:masterformula}
&\bigg[\begin{array}{lcr}
C_1(\mu_{Z}) \\
C_2(\mu_{Z})  \end{array}\bigg] \!\! = 
 e^{D(\mu_{Z})} P\exp\!\bigg(\!\int_{\mu_{m_{\chi}}}^{\mu_{Z}}\!\!\frac{\d\mu}{\mu} \hat\gamma \! \bigg) 
\bigg[\begin{array}{lcr}
C_1(\mu_{m_{\chi}}) \\
C_2(\mu_{m_{\chi}})  \end{array}\bigg].
\end{align}
This equation can be integrated analytically using $d\mu/\mu = d\alpha_2/\beta_2[\alpha_2]$.  For $X=ZZ,\gamma Z, \gamma\gamma$ we find
% for any neutral gauge boson final state: $X=ZZ,\gamma Z, \gamma\gamma$
\begin{align} \label{eq:sigmafactors}
\Sigma_1
 &=\frac{\e^{\Omega+D}}{3}\left(2\,z^{-\frac{4\psi}{b_0}}+z^{\frac{2\psi}{b_0}}\right),
\\
\Sigma_1-\Sigma_2 
  &=\frac{2\,\e^{\Omega+D}}{3}
  \left(z^{-\frac{4\psi}{b_0}}-z^{\frac{2\psi}{b_0}}\right),
  \nonumber 
\end{align}
where $\psi=1-i\pi$, $z=\alpha_2(\mu_{Z})/\alpha_2(\mu_{m_{\chi}})$, $D$ is in \eq{eq:lowscale2}, and $\Omega$ equals
\begin{eqnarray}
&&\Omega=\frac{-2\pi \Gamma_0^g\,\big(z\ln z\!+\!1\!-\! z\big)}{b_0^2\,\alpha_2(\mu_Z)}
   -\frac{\Gamma_0^g\, b_1 \big(\ln z \!-\! z\!-\! \frac{\ln^2 z}{2}\!+\! 1\big)}{2b_0^3}\nonumber\\
&&-\frac{\ln z}{2b_0}\bigg[8\Big(\ln\frac{4m_\chi^2}{\mu_{m_{\chi}}^2}-1\Big)-2b_0\bigg]-\frac{\Gamma_1^g}{2b_0^2}\left(z\!-\!\ln z\!-\! 1\right)\,.
\end{eqnarray}
Treating Sommerfeld effects at tree-level the ratio of cross sections is given by the Sudakov form factors
\begin{align}
&\frac{\sigma^{\text{NLL+\cancel{SE}}}_{\chi^+\chi^-\rightarrow X}}
 {\sigma^{\text{tree}}_{\chi^+\chi^-\rightarrow X}}
 = | \Sigma_1 |^2
  ,\qquad 
 \frac{\sigma^{\text{NLL+\cancel{SE}}}_{\chi^0\chi^0\rightarrow X}}
 {\sigma^{\text{tree}}_{\chi^+\chi^-\rightarrow X}}
  = | \Sigma_1 - \Sigma_2 |^2
 \,.\label{eq:delta12}
\end{align}
This nonzero result for $\chi^0\chi^0\to ZZ,Z\gamma,\gamma\gamma$ at short distances starts at NLL in $|\Sigma_1-\Sigma_2|^2$, and occurs because there is a Sudakov mixing between the $W^+W^-$ and $W^3W^3$ from soft gauge boson exchange. This is similar in spirit to the Sommerfeld mixing of the initial states. 

In Fig.~\ref{figNLL} we plot $|\Sigma_1|^2$ and $|\Sigma_1-\Sigma_2|^2$ as a function of $m_\chi$. To obtain theoretical uncertainty bands we use the residual scale dependence at LL and NLL obtained by varying $\mu_{m_\chi} = [m_\chi,4m_\chi]$ and $\mu_{Z} = [m_Z/2,2 m_Z]$.  The one-loop fixed order results of~\cite{Hryczuk:2011vi} are within our LL uncertainty band. Our NLL result yields precise theoretical results for these electroweak corrections.  To test our uncertainties we added non-logarithmic ${\cal O}(\alpha_2)$ corrections to $C_{1,2}(\mu_{m_\chi})$, of the size found in~\cite{Hryczuk:2011vi}, and noted that the shift is within our NLL uncertainty bands.

{\bf Indirect Detection Phenomenology}
Combining Eqs.~\ref{eq:sommerfeldxsec} and~\ref{eq:sigmafactors} with the standard Sommerfeld enhancement (SE) factors $s_{00}$ and $s_{0\pm}$, we can now compute the total cross section for annihilation to line photons at NLL+SE and compare to existing limits from indirect detection. We sum the rates of photon production from $\chi^0 \chi^0 \rightarrow \gamma \gamma, \gamma Z$, as the energy resolution of current instruments is typically comparable to or larger than the spacing between the lines (see e.g. \cite{Cohen:2013ama} for a discussion). 

In Fig.~\ref{figxsec} we display our results for the line cross sections calculated at LL+SE and NLL+SE. Our theoretical uncertainties are from $\mu_{m_\chi}$ variation. (The $\mu_{Z}$ variations are very similar. Since both cases are dominated by the variation of the ratio of the high and low scales we do not add them together.) In the left panel we compare to earlier cross section calculations, including ``Tree-level + SE'' where Sudakov corrections are neglected, the ``One-loop fixed-order'' cross section where neither Sommerfeld or Sudakov effects are resummed (taken from \cite{Fan:2013faa}), and the calculation in~\cite{Hryczuk:2011vi} where Sommerfeld effects are resummed but other corrections are at one-loop. At low masses, our results converge to the known ones (except~\cite{Hryczuk:2011vi} which focused on high masses and omits a term that becomes leading-order at low masses). At high masses, our NLL+SE result provides a sharp prediction for the annihilation cross section with $\simeq 5\%$ theoretical uncertainty.

In the right panel of Fig.~\ref{figxsec} we compare the NLL cross section to existing limits from H.E.S.S \cite{Abramowski:2013ax} and projected ones from CTA. In the latter case we follow the prescription of \cite{Cohen:2013ama}, based on \cite{Bergstrom:2012vd}, and in both cases we assume an NFW profile with local DM density 0.4 GeV/cm$^3$. We assume here that the $\chi^0$ constitutes all the DM due to a non-thermal history (the limits can be straightforwardly rescaled if it constitutes a subdominant fraction of the total DM). For this profile, we see that H.E.S.S already constrains models of this type for masses below $\sim 4$ TeV, consistent with the results of \cite{Cohen:2013ama} (which employed the tree-level+SE approximation), and that five hours of observation with CTA could extend this bound to $\sim 10$ TeV.
Any constraint on the line cross section should be viewed as a joint constraint on the fundamental physics of DM and the distribution of DM in the Milky Way
%~\cite{*[{}] [{Assuming an Einasto profile with standard parameters ($\alpha=0.17$, $r_s=20$ kpc) would strengthen the limits by roughly a factor of 2; allowing a Burkert profile with a core radius greater than $\sim 2$kpc would weaken the limits (albeit such a large core is disfavored by simulations), by a factor of $\sim 20$ for a 5kpc core and $\sim 100$ for a 10kpc core.}] profile}.
\footnote{Assuming an Einasto profile with standard parameters ($\alpha=0.17$, $r_s=20$ kpc) would strengthen the limits by roughly a factor of 2; allowing a Burkert profile with a core radius greater than $\sim 2$kpc would weaken the limits, e.g. by a factor of $\sim 20$ for a 5kpc core and $\sim 100$ for a 10kpc core.}.

The method we developed here allows systematically improvable effective field theory techniques to be applied to DM, and enabled us to obtain NLL+SE predictions for the DM annihilation cross section to photon lines. This enables precision constraints to be placed on DM.

{\it{Note added:}}\ 
As our paper was being finalized two papers appeared \cite{Baumgart:2014vma,BauerCohenHillSolon} which also investigate DM with SCET. They are complementary to ours:~\cite{Baumgart:2014vma} computes the semi-inclusive cross-section for fermionic DM annihilation at LL, and~\cite{BauerCohenHillSolon} investigates the exclusive line annihilation cross section for scalar DM up to NLL.

%\vspace{9mm}
{\it Acknowledgments:} 
This work is supported by the U.S. Department of Energy under grants DE-SC00012567 and DE-SC0011090, and by the Simons Foundation Investigator grant 327942 to IS. TS thanks Timothy Cohen for discussions.

%\bibliographystyle{doiplain}
%   \let\oldnewblock=\newblock
%    \newcommand\dispatcholdnewblock[1]{\oldnewblock{#1}}
 %   \renewcommand\newblock{\spaceskip=0.3emplus0.3emminus0.2em\relax
%                         \xspaceskip=0.3emplus0.6emminus0.1em\relax
 %                        \hskip0ptplus0.5emminus0.2em\relax
 %                      {\catcode`\.=\active
%                    \expandafter}\dispatcholdnewblock}
%\bibliography{darkmatter}

\bibliographystyle{h-physrev}
\bibliography{bibliography}

\end{document}